\documentclass[preprint,3p,twocolumn,english,number,times]{elsarticle}
\biboptions{sort&compress}
\usepackage{graphicx}
\usepackage{babel}
\usepackage[pdftex,
            colorlinks=true,
            hyperindex,
            plainpages=false,
            pagebackref=true,
            bookmarksopen,
            bookmarksnumbered
]{hyperref}
\graphicspath{{./figures/}}
\usepackage{amssymb,amsmath}

\begin{document}

\author[temple]{S.~Joosten}
\ead{sylvester.joosten@temple.edu}
\author[temple]{E.~Kaczanowicz}
\ead{edkacz@temple.edu}
\author[JLab]{M.~Ungaro}
\ead{ungaro@jlab.org}
\author[temple]{M.~Rehfuss}
\author[temple]{K.~Johnston}
\author[temple]{Z.-E.~Meziani}
\ead{meziani@temple.edu}
\address[temple]{Temple University, Department of Physics (035-08), 1925 N. 12th Street, Philadelphia, PA 19122-1801}
\address[JLab]{Thomas Jefferson National Accelerator Facility, 12000 Jefferson Avenue, Newport News, VA 23606}

\title{\large Enhanced UV light detection using a p-terphenyl wavelength shifter}

\begin{abstract}
  UV-glass photomultiplier tubes (PMTs) have poor photon detection efficiency
  for wavelengths below $300\,\text{nm}$ due to the opaqueness of the window
    material.  Costly quartz PMTs could be used to enhance the efficiency below
    $300\,\text{nm}$.  A less expensive solution that dramatically improves this
    efficiency is the application of a thin film of a p-terphenyl (PT)
    wavelength shifter on UV-glass PMTs.  This improvement was quantified for
    Photonis XP4500B PMTs  for wavelengths between $200\,\text{nm}$ and
    $400\,\text{nm}$.  The gain factor ranges up to 5.4 $\pm$ 0.5 at a
    wavelength of $215\,\text{nm}$, with a material load of
    $110\pm10\,\mu\text{g}/\text{cm}^2$ ($894\,\text{nm}$).  The wavelength
    shifter was found to be fully transparent for wavelengths greater than
    $300\,\text{nm}$.  The resulting gain in detection efficiency, when used in
    a typical \u{C}erenkov counter, was estimated to be of the order of 40\%.
    Consistent coating quality was assured by a rapid gain testing procedure
    using narrow-band UV LEDs.  Based on these results, 200 Photonis XP4500B
    PMTs were treated with PT for the upgraded low-threshold \u{C}erenkov
    counter (LTCC) to be used in the CEBAF Large Acceptance Spectrometer
    upgraded detector (CLAS12) at the Thomas Jefferson National Accelerator
    Facility.
\end{abstract}
\maketitle

\section{Introduction}

During the 6 GeV maximum electron beam energy era of Jefferson Lab, a
\u{C}erenkov Counter~\cite{ltcc:2001nim} (CC) has been used in several
experiments in Hall B for electron/pion discrimination as part of the CLAS
detector data acquisition trigger and particle identification systems.
The CC has been refurbished to be used as a LTCC for pion/kaon discrimination in
the new detector~\cite{CLAS12:2008tr} for the $12\,\text{GeV}$ upgrade of
Jefferson Lab.
Pions in the momentum region of interest produce significantly less \u{C}erenkov
photons compared to electrons.
Hence, a significant increase in the light-collection efficiency was
essential in the CC upgrade.

The number density of \u{C}erenkov photons with a wavelength $\lambda$ for a
particle with velocity $v$ and charge $ze$ passing through a unit of length of a
radiator can be written as~\cite{Leo:1987kd},

\begin{align}
\frac{d^2n}{dxd\lambda} = \frac{2\pi z^2\alpha}{\lambda^2}\sin^2{\theta_C(v)},
\end{align}

with the fine structure constant $\alpha$ and the \u{C}erenkov cone half-angle
$\theta_C$.
Due to the $\lambda^{-2}$ dependence of this distribution, 
a good sensitivity to photons in the UV region is required 
in order to maximize the efficiency of a \u{C}erenkov detector.

The detector efficiency $\eta_C(\lambda)$ for a \u{C}erenkov counter is
proportional to the transparency of the \u{C}erenkov medium $\eta_T$, the
reflectivity of the mirrors $\eta_R$ and the quantum efficiency (QE) of the
photomultipliers (PMTs), defined as the ratio of the number
of photoelectrons emitted by the photocathode to the number of photons incident on
the window,  $\eta_P$,
\begin{align}
  \eta_C(\lambda) \propto \eta_T(\lambda)\eta_R(\lambda)\eta_P(\lambda).
\end{align}
For a typical \u{C}erenkov detector, the medium stops being transparent for light
below $200\,\text{nm}$ to below $150\,\text{nm}$, depending on the medium used,
while mirror reflectivity commonly drops off below approximately $180\,\text{nm}$.
For PMTs with photocathodes of the bialkali family, the limiting factor of
enhanced QE in the UV region is the transparency of the window
material~\cite{Photonis:2007ta}.
Borosilicate windows can be used to detect wavelengths as low as approximately
$300\,\text{nm}$, UV-glass windows
down to $250\,\text{nm}$, and only quartz windows can efficiently detect light
down to $180\,\text{nm}$.

While quartz windows maximize the UV-sensitivity of the PMT, and therefore the
\u{C}erenkov detector performance, they are difficult and expensive to produce. 
In fact, most newer types of PMTs are only available with borosilicate and
UV-glass windows\footnote{For example the H8500,
H10966 and H12700 generations of multi-anode PMTs produced by Hamamatsu are
only available with borosilicate and UV-glass windows.}.

A wavelength shifter (WLS) deposited on the face of a borosilicate or UV glass
PMT provides an effective alternative to boost the efficiency of a \u{C}erenkov
detector by converting UV photons with a wavelength below $300\,\text{nm}$ into two
isotropically emitted photons with longer wavelengths.
Additional requirements for a WLS material are a short decay time, a high
transparency to its own emitted wavelength as well as longer wavelengths, and
stability against evaporation and aging.
Examples of such WLS materials include p-terphenyl (PT),
p-quaterphenyl (PQ), tetraphenyl-butadiene (TPB) and diphenylstilbene
(DPS)~\cite{Mai:71,Garwin:1973ex,Alves:1974em,Koczon:1457653}.
PT in particular has been shown to provide the highest gain in photodetection
efficiency for PMTs of the bialkali family as its emission spectrum matches the
region of peak performance of the bialkali photocathode while having a short
decay time near
$1\,\text{ns}$~\cite{Garwin:1973ex,Baillon:1975jx,Eigen:1979ev,Grande1983539,Gorin:1986ej,Koczon:1457653}.

\section{The CLAS \u{C}erenkov Counter upgrade}

\begin{figure}
\centering
\includegraphics[width=1\linewidth]{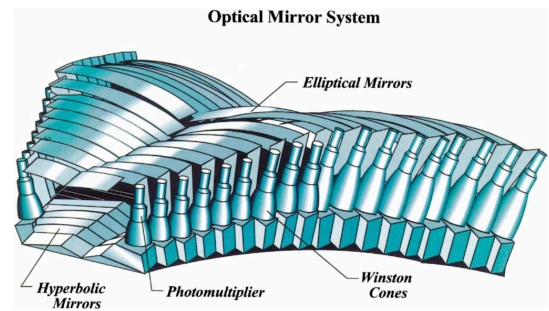}
\includegraphics[width=1\linewidth]{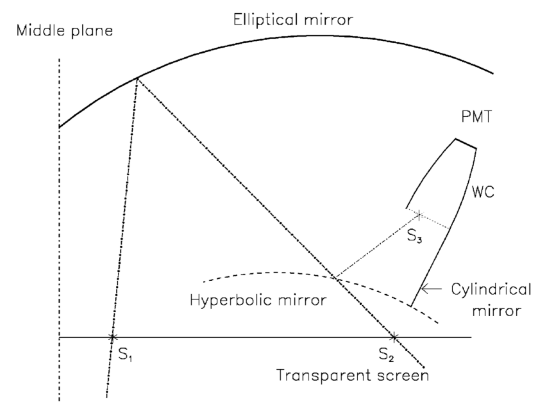}
\caption{
  The CLAS CC. Top: one sector consist of 6 elliptical and
  36 hyperbolic mirrors focusing the \u{C}erenkov light onto 36 PMTs. Bottom:
  the light is reflected and focused by the mirror systems onto the Winston
  Cones and the PMTs.
\label{fig:ltcc-clas}}
\end{figure}

Application of a WLS coating on UV-glass PMTs can enable a refurbished
\u{C}erenkov detector to hit key performance goals in a more demanding
experimental environment compared with the original detector.
In fact, this was crucial for the upgrade of the CLAS CC to be used as the
CLAS12 LTCC.

The CC consists of six sectors, each equipped with
36 elliptical and 36 hyperbolic mirrors focusing the \u{C}erenkov light onto 36
PMTs, as shown in Fig.~\ref{fig:ltcc-clas}. 
The sectors are filled with $C_4F_{10}$ gas, with a refraction index of
$1.0014$.
Electrons cross through approximately $1\,\text{m}$ of gas, while the produced
\u{C}erenkov photons undergo between 3 and 4 bounces before they reach the
PMT face.
The average resulting light yield at the PMT face in the original CC was
observed to be 20 photons.
Without WLS, a small fraction of these photons are detected due to a poor PMT
efficiency for UV light.

\begin{figure}
\centering
\includegraphics[width=1\linewidth]{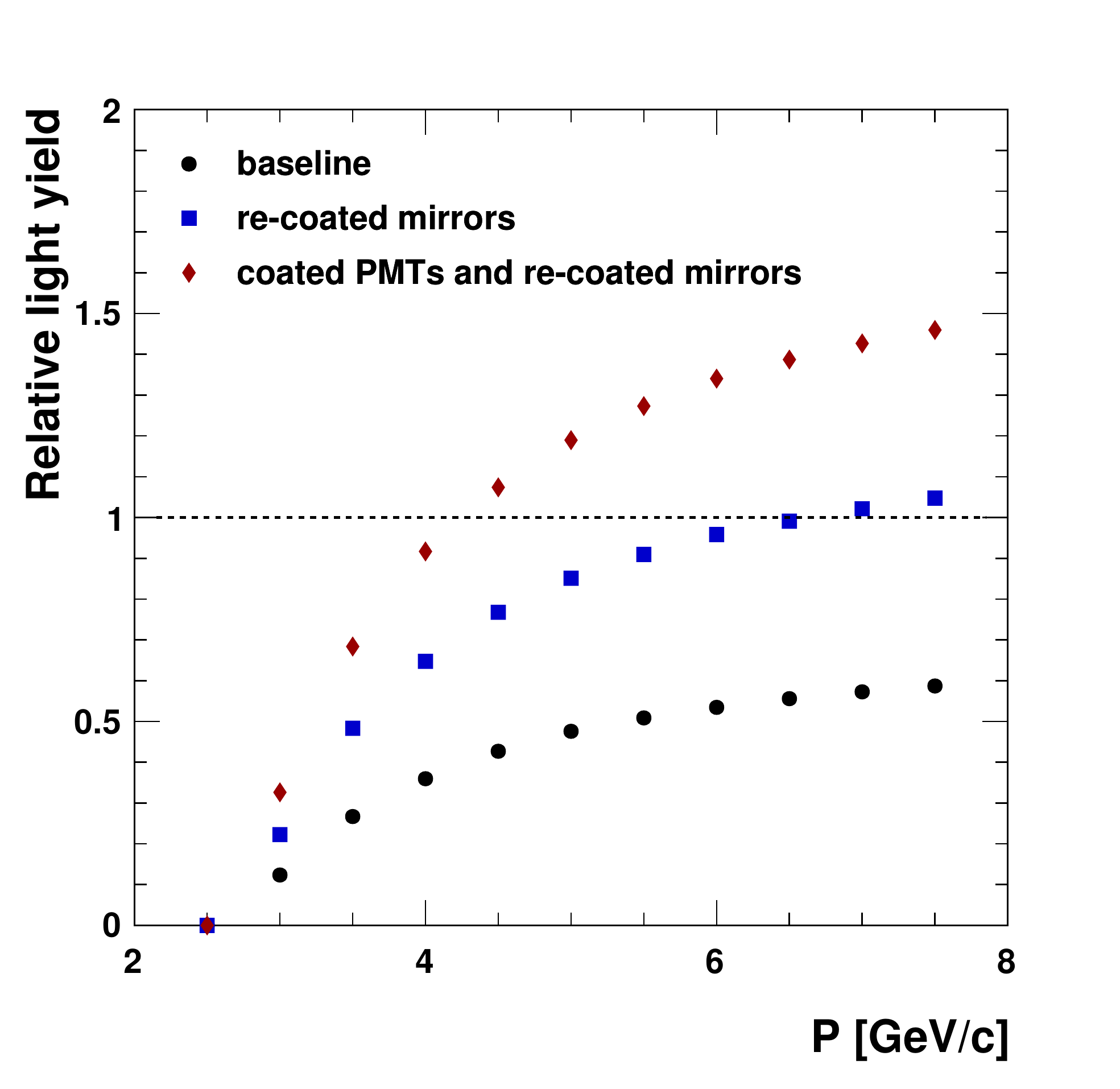}
\caption{
  The pion light yield at the PMT face in the LTCC, relative to the electron
  light yield in the baseline CC. 
  The pion light yield with only the increase in gas volume (black circles), 
  and the addition of a refurbished mirror system (blue squares) is
  too low to be used for pion identification. The PT coating on the PMTs (red
  diamonds) is necessary to achieve an acceptable light yield in the kinematic
  region of interest.
\label{fig:ltcc-nphe}
}
\end{figure}

The pion \u{C}erenkov  threshold in $C_4F_{10}$ is $2.6\,\text{GeV}/c$, while
the kaon threshold is $9.4\,\text{GeV}/c$.
Hence, after a track is found not to be an electron or proton by the rest of the
particle-identification system in CLAS12, it can be further identified as either
a pion or kaon by the LTCC.
To perform this pion/kaon separation function with a good efficiency, it is
crucial that pions generate a sufficient number of \u{C}erenkov photons.
It was found that pions in the momentum region of interest produce less than
half the number of \u{C}erenkov photons compared to electrons, and the detector
efficiency drops rapidly with a decreasing number of \u{C}erenkov photons.
In order to maximize the LTCC light yield, the CC gas volume has been increased
by $20\%$, the mirror and Winston cone system has been refurbished with an
improved reflective coating\footnote{The elliptical and hyperbolic mirrors were
covered with aluminized Lexan strips, and the cylindrical mirrors and Winston
cones were re-coated with $\text{Al}+\text{MgF}_2$ by Evaporated Coating Inc.
(ECI).},
and the PMTs have been coated with a PT wavelength shifter.
The results of this detector upgrade are shown in Fig.~\ref{fig:ltcc-nphe}.
The PT coating is necessary to ensure a pion light yield in the LTCC that is
comparable (or better) then the original electron light yield in the CC.
This increased light yield is required in order to reliably separate pions
from kaons with the LTCC.

\section{Wavelength Shifter Thickness}

\begin{figure}
\centering
\includegraphics[width=1\linewidth]{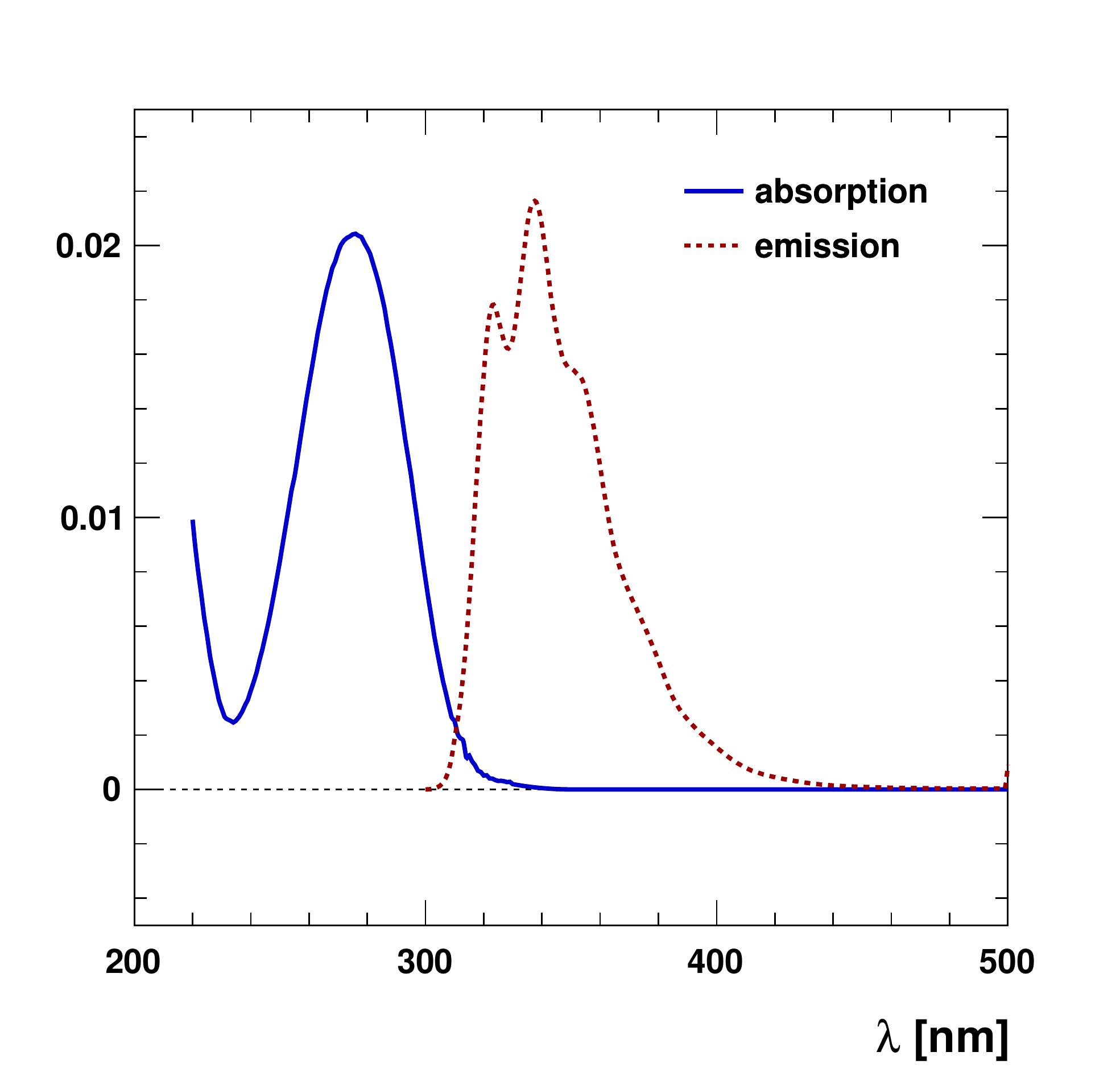}
\caption{
Absorption (solid blue line) and emission spectra (dashed red line) of
PT in arbitrary units (normalized to unity). Data from \cite{Dixon:2005cp}.
\label{PT-spectra}}
\end{figure}

\begin{figure}
\centering
\includegraphics[width=1\linewidth]{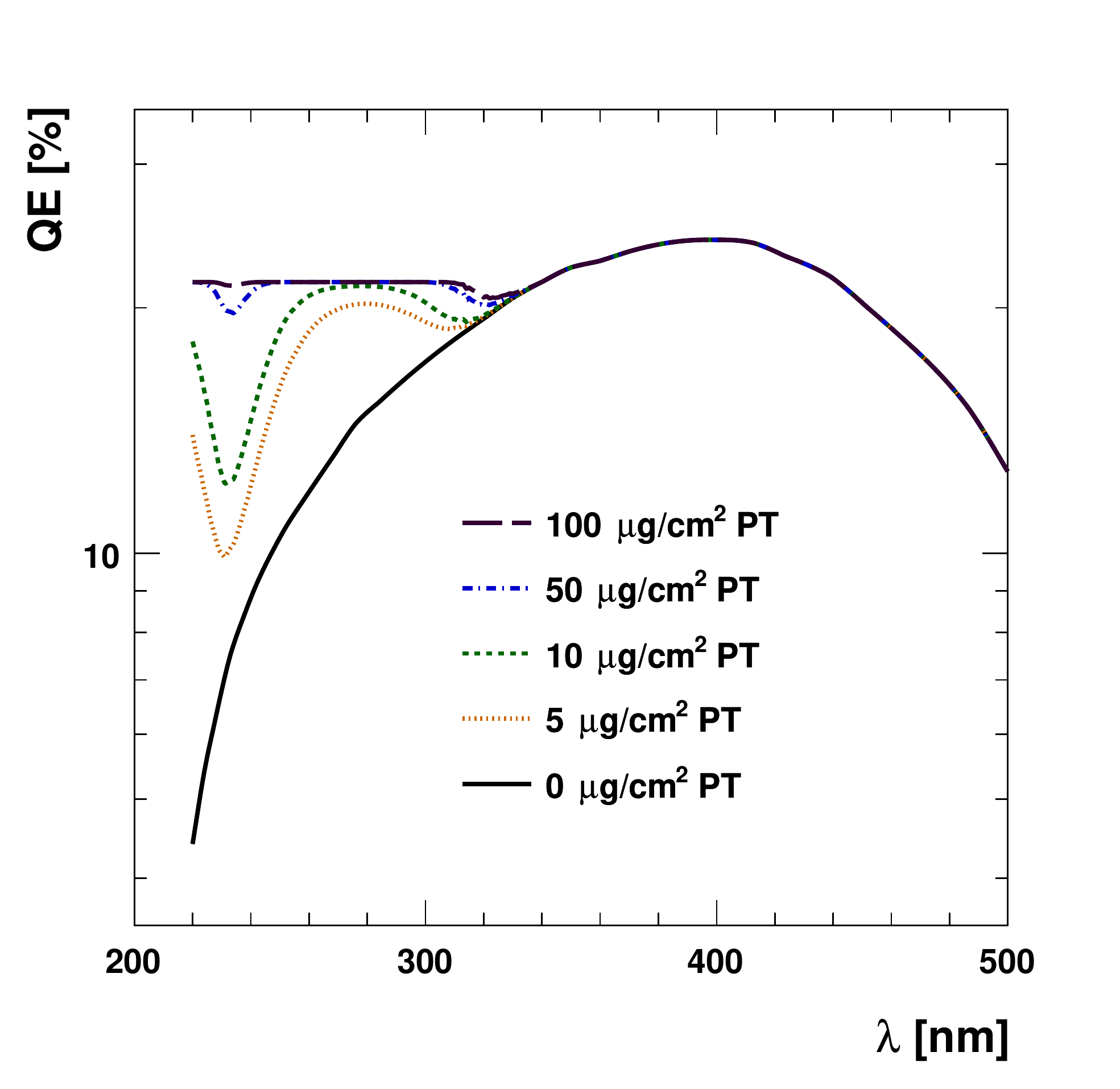}
\includegraphics[width=1\linewidth]{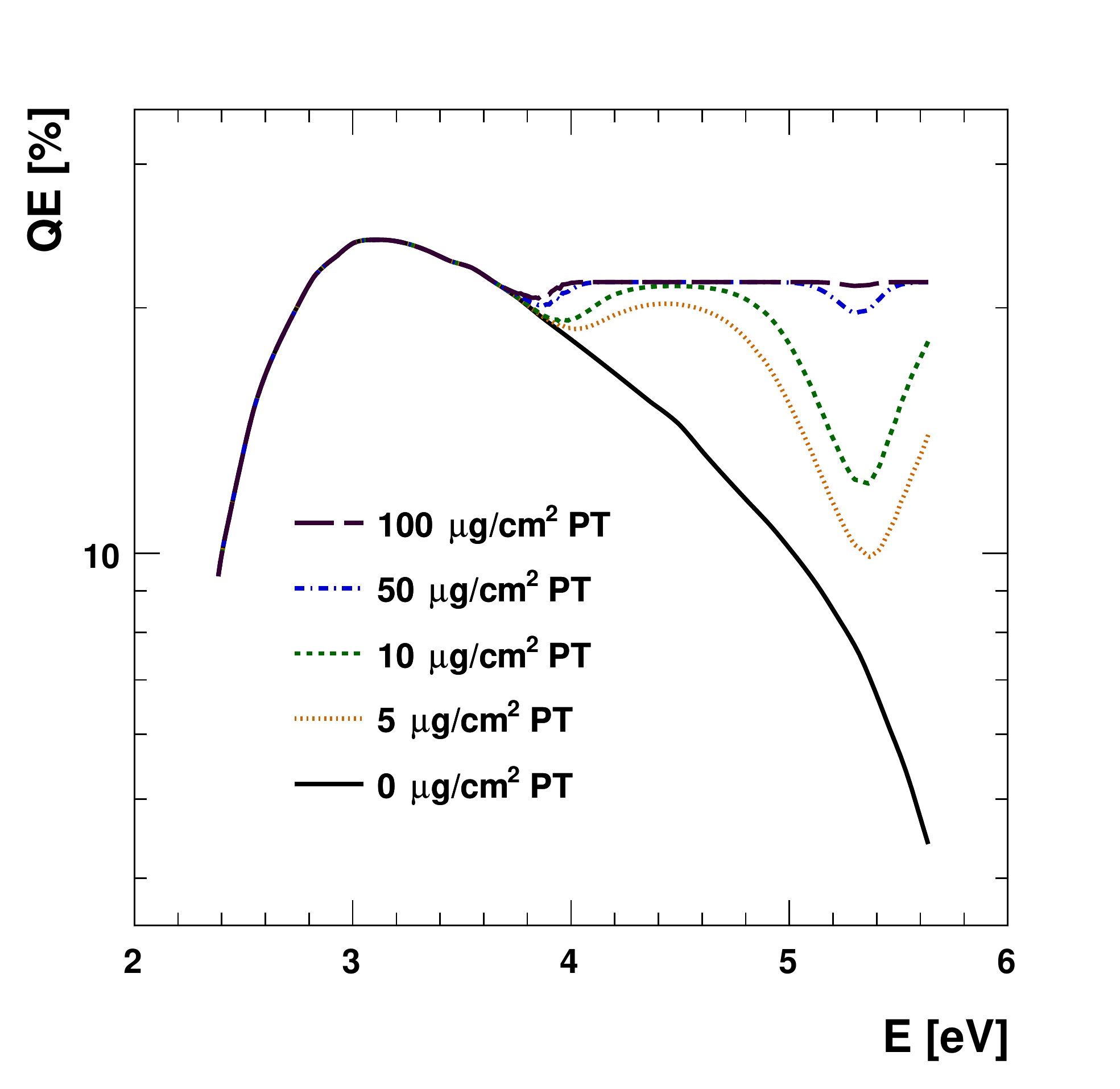}
\caption{
The typical QE for a Photonis XP4500B PMT (solid black line), compared to the
projected QE after application of a PT wavelength shifter at four
different material loads (dashed lines), as a function of wavelength (top) and
energy (bottom).
\label{thqe}}
\end{figure}

\begin{figure}
\centering
\includegraphics[width=1\linewidth]{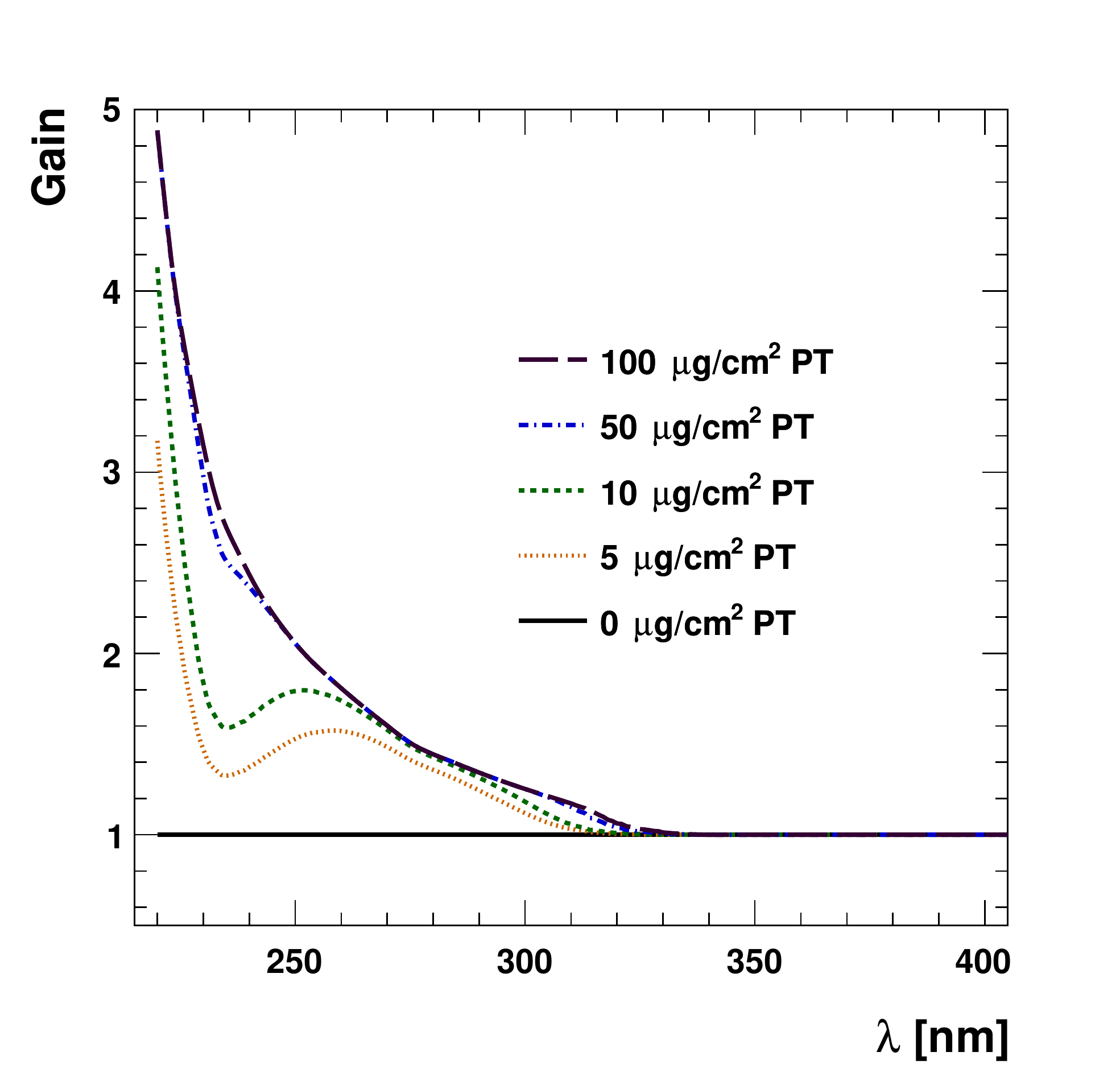}
\caption{
  Nominal prediction of the gain in QE for a typical Photonis XP4500B PMT at various
  material loads of a PT wavelength shifter.
  \label{thgain}
}
\end{figure}

The ideal thickness of the PT film is determined by the minimum thickness
required to ensure a good absorption in the WLS. Thicker layers will lead to
a decline in optical transparency above the absorption region, as well as an
increase in background due to scintillations excited directly by stray fast
particles
\cite{Garwin:1973ex}.
The ideal minimum thickness can be estimated by calculating 
the expected QE after application of the wavelength shifter film
with thickness $\Delta$,
\begin{align}
\eta_\text{WLS}(\lambda, \Delta)
&=
(1-\mathcal{P}_\text{abs}(\lambda, \Delta))\eta_P(\lambda) \nonumber\\
 &\quad+ \mathcal{P}_\text{abs}(\lambda, \Delta)
      \int d\lambda^\prime
      \eta_P(\lambda^\prime)
      I_\text{ems}(\lambda^\prime)Q_f(\lambda^\prime),
\end{align}
with $\mathcal{P}_\text{abs}(\lambda, \Delta)$ the probability to absorb a
photon with wavelength $\lambda$ in a WLS with thickness $\Delta$ and
$I_\text{ems}(\lambda^\prime)$ the distribution of emitted photons incident on the
PMT window, and $Q_f(\lambda^\prime)$ the QE of the WLS.
In this paper, we consider the UV-glass Photonis XP4500B PMT, using the QE data
from \cite{Photonis:2007ta}, 
PT absorption and emission data from \cite{Dixon:2005cp} (shown in
Fig.~\ref{PT-spectra},
a density of 1.23 $\text{g}/\text{cm}^3$ and a molecular weight of 230.30376
g/mol.
The results are shown in Fig.~\ref{thqe} and a nominal prediction of the gain
($\eta_\text{WLS}/\eta_P$) in Fig.~\ref{thgain}.
The absorption is starting to saturate at a material load of 50 $\mu
\text{g}/\text{cm}^2$, and around 100 $\mu \text{g}/\text{cm}^2$ a near-optimal level of
saturation is reached. The potential increase in gain for higher material loads
is minimal. The WLS optical transparency remains very good up to $300\,\mu
\text{g}/\text{cm}^2$, resulting in a wide
plateau region between $100$-$300\,\mu
\text{g}/\text{cm}^2$~\cite{Garwin:1973ex,Baillon:1975jx,Koczon:1457653}. Due to
this wide plateau region, the impact of small fluctuation in coating thickness
is negligible.

\section{Coating Procedure}
PT deposition has been accomplished in the past through vacuum
evaporation~\cite{Garwin:1973ex,Baillon:1975jx,Koczon:1457653} or
alternatively by means of a solution with a plastic binder in an organic 
solvent~\cite{Eigen:1979ev,Grande1983539,Gorin:1986ej}, both with varying
degrees of success.

\begin{figure}
\includegraphics[width=1\linewidth]{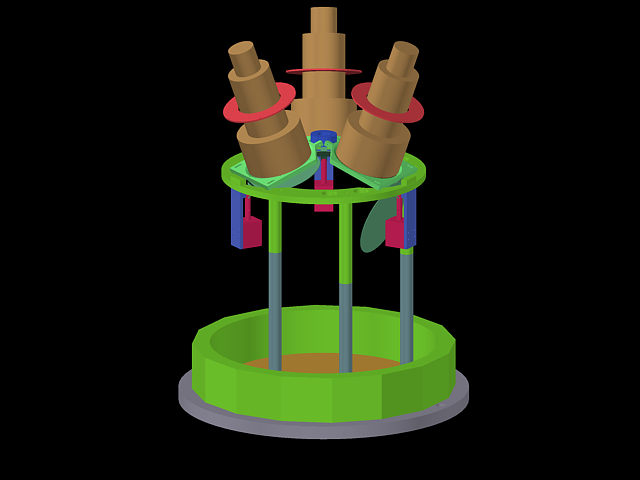}
\includegraphics[width=1\linewidth]{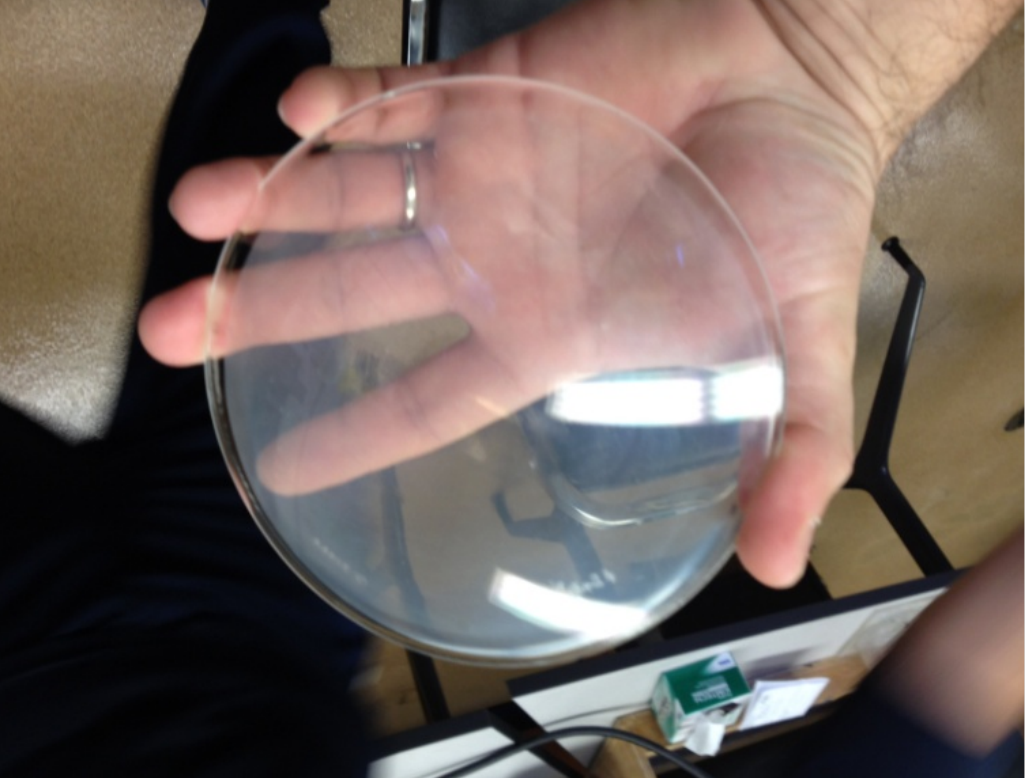}
\includegraphics[width=1\linewidth]{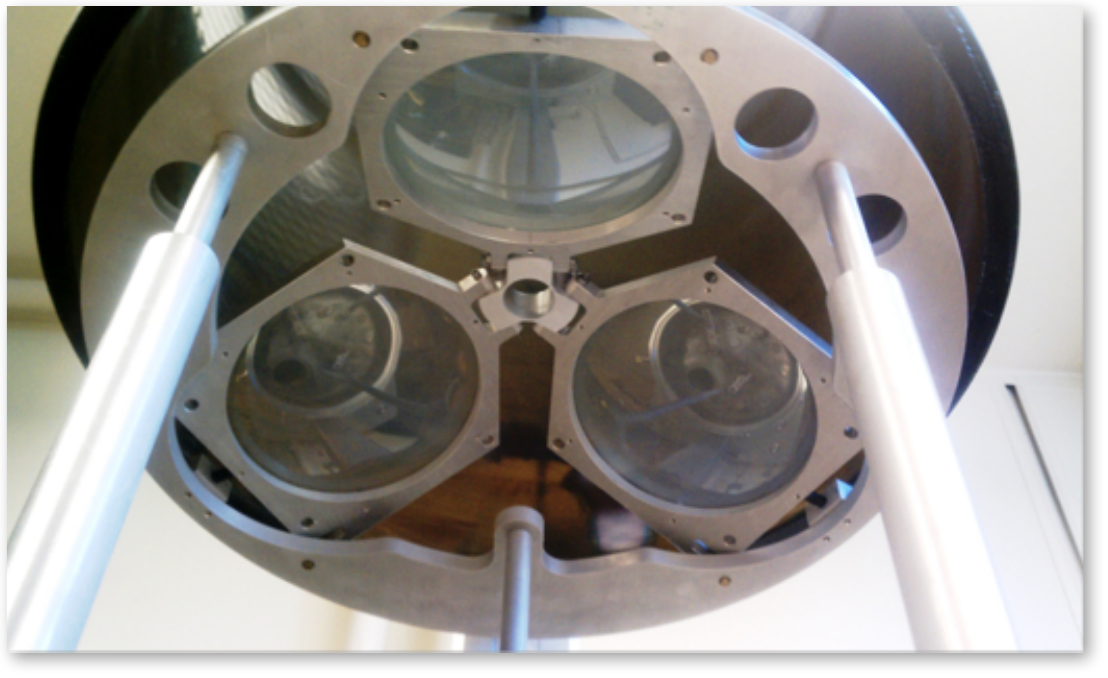}
\caption{
Top: Schematic overview of the PMT carousel holding three 5-inch Photonis XP4500B
PMTs. The carousel is located directly inside the vacuum evaporator.
Center: a test glass after coating. 
Notice that  $110\,\mu\text{g}/\text{cm}^2$ ($894\,\text{nm}$) of PT provides an
almost transparent coating.
Bottom: picture of the actual setup.
\label{fig:carousel}}
\end{figure}

For this work, the PT WLS\footnote{p-Terphenyl powder with 99+\% purity.} was deposited directly on the UV-glass
window of the Photonis XP4500B PMTs in a vacuum evaporator, which allows for the
coating thickness to be precisely controlled.
The setup is capable of coating three 5-inch PMTs simultaneously, as illustrated
in Fig.~\ref{fig:carousel}.

\subsection{Cleaning the PMT}
To obtain a homogeneous coated PMT it is crucial that all impurities are removed
from the PMT surface prior to coating.
To accomplish this the PMT face is cleaned off using isopropanol and then
immediately dried with a blast of compressed dry air or nitrogen.
This procedure is repeated 3 to 5 times as needed.
If the PMT was previously coated, this coating is first carefully removed with 3
to 5 cleaning cycles with acetone.

\subsection{Evacuation phase}

Three PMTs can be simultaneously loaded into the vacuum evaporator (see
Fig.~\ref{fig:carousel} top, bottom). 
The PT powder is loaded into a $0.75\,\text{cm}^3$ modified resistive heated
boat at the center of the base of the evaporator.
This boat is covered with a 200 mesh screen to prevent the PT powder from spilling.
The air is evacuated using a mechanical pump and a turbo pump.
The chamber is ready for deposition once vacuum reaches an absolute pressure of
$5\times10^{-6}\,\text{Torr}$.
This evacuation phase takes approximately 12 hours in our evaporator.

\subsection{PT deposition}
The boat is heated using a DC current of approximately $25\,\text{A}$ for a
temperature of $150^\circ\text{C}$,
initiating the sublimation of the PT powder. 
A water-cooled quartz deposition monitor, 
mounted at the same radial distance from the boat as the PMT
face, continuously measures the evaporation rate. The precise current is
controlled to ensure a deposition rate of  $1.5\,\text{nm}/\text{s}$.
The PMTs remain shuttered for approximately 6 minutes, until this target
evaporation rate has been reached. 
Once evaporation conditions are optimal, the shutters are opened exposing the
PMT faces to the evaporant.
When the target thickness is reached, the shutters are closed and the power to
the boat is shut off.

\subsection{Material load calibration}

To precisely calibrate the material load on the PMT face, a series of
evaporations was performed on a thin glass surface ("watch glass") with an
identical radius and curvature as the Photonis XP4500B PMT face. By weighing the
before and after deposition, the exact material load on the PMT face could be
linked to the amount of PMT powder in the evaporation boat. It was found that,
for our setup, an optimal material load of $110\,\mu\text{g}/\text{cm}^2$
($894\,\text{nm}$) is obtained with $0.85\pm0.01\,\text{g}$ of PT.

\section{Wavelength Dependent Gain Measurement}

The wavelength dependent gain in PMT efficiency was measured using a
monochromator (Newport model CS260-USB-1-FH-A) with a deuterium light source 
with a reach between $200\,\text{nm}$ and $400\,\text{nm}$. 
The sections below discuss the exact experimental setup as well as the measured
results.

\subsection{Experimental setup}
\begin{figure}
\centering
\includegraphics[width=1\linewidth]{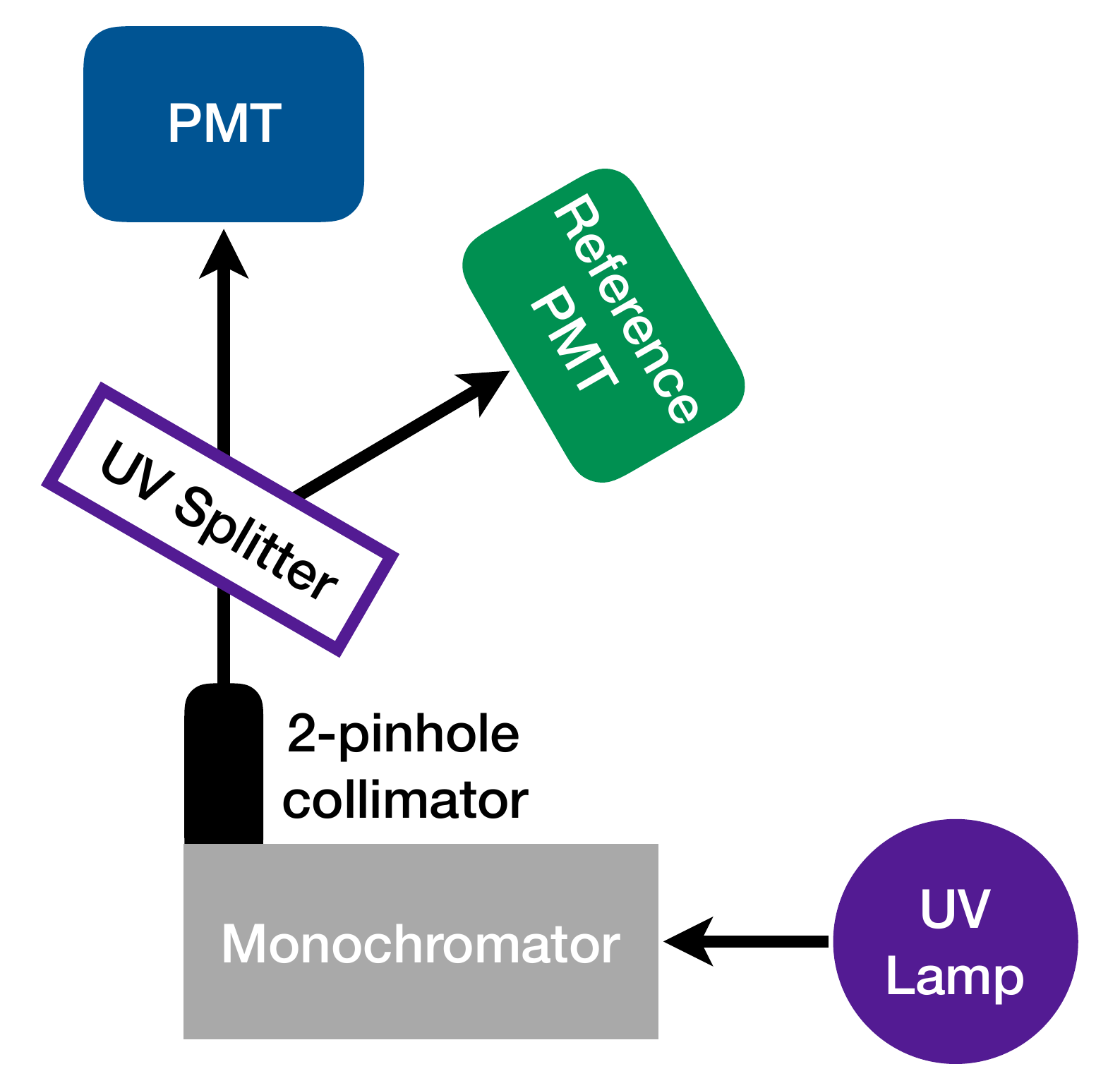}
\caption{
  Monochromator setup used for the gain measurement.
\label{fig:monoChroSetup}}
\end{figure}

The deuterium light source fully covers the
spectrum between $200\,\text{nm}$ and $400\,\text{nm}$. The wavelength can be
selected by the monochromator with a precision of 
$0.35\,\text{nm}$\cite{Newport::CS260-USB-1-FH-A}. 
The light beam of the selected wavelength is collimated using two
$760\,\mu\text{m}$ pinholes. 
The beam is then split with a UV Grade fused silica beam splitter in two separate beams, 
as shown in Fig.~\ref{fig:monoChroSetup}.
Each light beam illuminates the face of a single PMT: the PMT to
be tested, and a calibration PMT. The beam at the face of the PMT consists
of single photons at a rate of $30\,\text{kHz}$ to $3\,\text{MHz}$ depending on
the wavelength. Note that the monochromator is not placed in a vacuum, and therefore
shorter wavelengths (close to $200\,\text{nm}$) undergo losses due
to absorption by the air.

The observed photon rate was read out by a scaler with a $-24\,\text{mV}$
threshold. Each PMT was measured while coated and then again after the coating
had been removed using acetone and isopropanol. The measurement was performed on
two separate PMTs. For each PMT, the measurement was performed in the standard
configuration and in a configuration where the calibration PMT and the test PMT
position were switched.
In all cases, the calibration PMT rates were found to be stable between
measurements.

The background was determined by an additional measurement with the light source
switched off. This background rate is subtracted from the observed rate.

\subsection{Monochromator Results}

\begin{figure}
\centering
\includegraphics[width=1\linewidth]{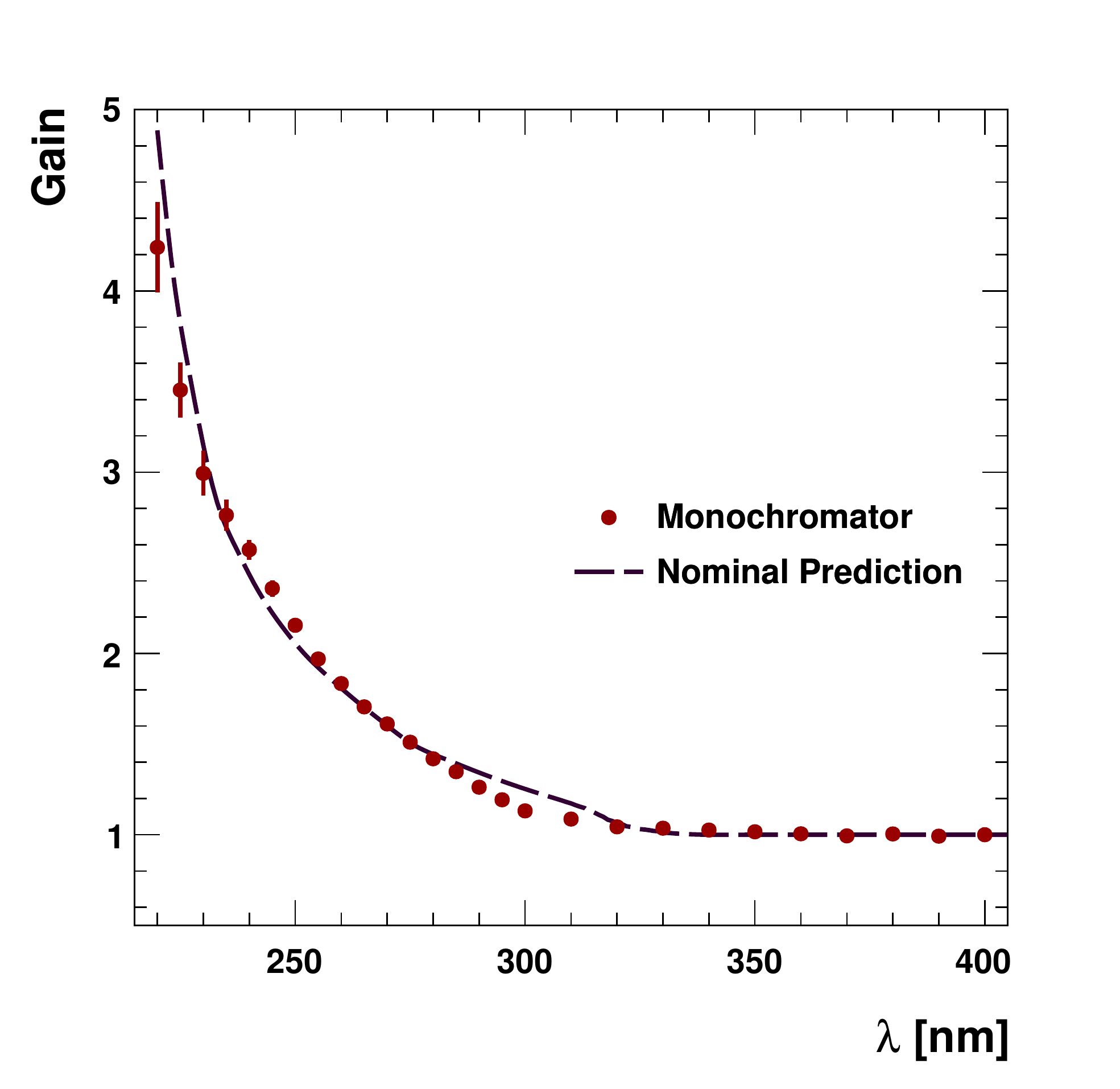}
\caption{
  Results from the monochromator measurement (red circles), compared to the
  nominal prediction for a PT material load of
  $110\,\mu\text{g}/\text{cm}^2$ (dashed line).
\label{fig:comp}}
\end{figure}

The results from 2 different PMTs each at 2 different positions were found to be
consistent. Fig.~\ref{fig:comp} shows the average measured gain from these four
measurements. The observed gain was found to be consistent with the nominal
prediction for a perfectly saturated coating.
Note that these results do not show any signs of a decline in WLS gain below
$240\,\text{nm}$, in contrast with the previous results from
Refs.~\cite{Eigen:1979ev,Gorin:1986ej}, consistent with the more recent
results from Ref.~\cite{Koczon:1457653}.

For a typical light-gas \u{C}erenkov detector, this gain in PMT efficiency
corresponds to an increase of approximately 40\% in total detector efficiency.

\section{Rapid Quality Assurance with UV LEDs}
The monochromator test procedure is too time-consuming to perform on all 200
PMTs for the CLAS12 LTCC.
For this reason, we implemented a rapid quality assurance (QA) procedure using
two UVTOP LEDs\footnote{Produced by Sensor Electronic Technology, Inc. (SETi)}
to ensure consistent results. 
The QA testing can be performed with up to 3 PMTs simultaneously.

\subsection{Procedure}
The procedure consists of three steps: gain matching, baseline test,
and assessment of the improved response.

To select the proper high-voltage settings for the PMTs, they were gain-matched
using the single photo-electron (SPE) spectrum, extracted
from the dark current when self-triggering on $-4\,\text{mV}$. This
step was also used to identify bad-performing PMTs.

For the next two steps, we used two UVTOP LEDs, at $265\,\text{nm}$ (UVTOP260)
and $290\,\text{nm}$ (UVTOP285), both with a line width of $6\,\text{nm}$.
Driven by an Agilent 33522A Function/Arbitrary Waveform Generator, they emit short
sub-nanosecond bursts of light. An average burst contains approximately 100
photons. The PMTs are read out by a CAEN V1729A flash ADC (FADC) triggered by
the function generator. The average number of photons converted in the
photocathode can be estimated by fitting the integrated ADC spectrum with a
Gaussian ansatz \cite{Bellamy:1994bv},

\begin{align}
  N_\text{PE} \approx \frac{\mu^2}{\sigma^2},
\end{align}
with $\mu$ and $\sigma$ the mean and with of the Gaussian.

\begin{figure}
\centering
\includegraphics[width=1\linewidth]{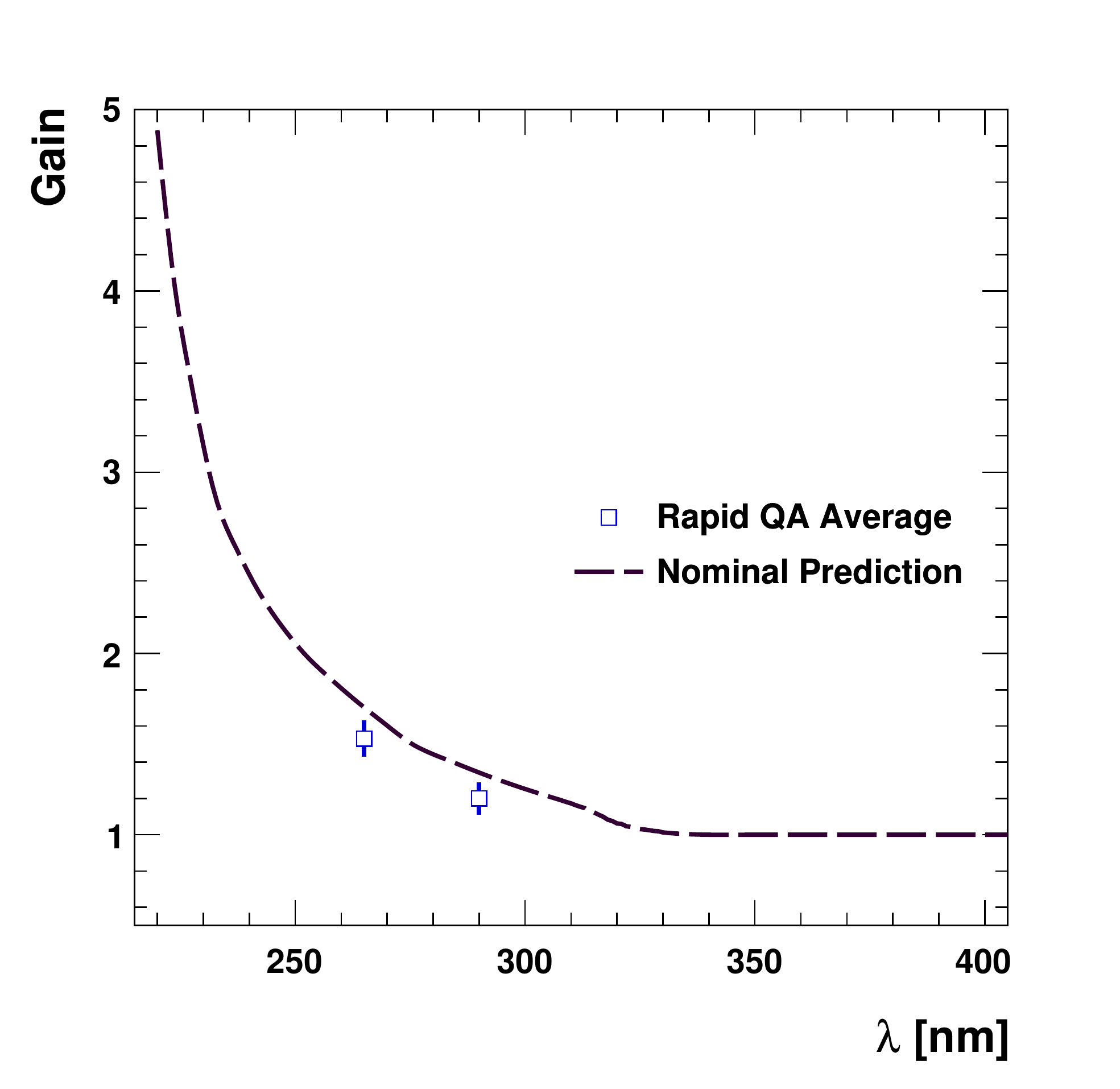}
\caption{
The average gain from the 200 coated XP4500B PMTs measured using the calibrated
2-LED rapid QA setup (blue squares), compared to the 
  nominal prediction for a PT material load of
  $110\,\mu\text{g}/\text{cm}^2$ (dashed line).
\label{fig:qa-comp}}
\end{figure}

This procedure is repeated after the PMT face has been coated.
To control for potential variations in measurement conditions before and after
coating, we use a Photonis XP4318B 3-inch PMT with a quartz entrance window as a
fourth calibration PMT.  
The average measured gain from the 200 XP4500B PMTs coated
for the CLAS12 LTCC is shown in Fig.~\ref{fig:qa-comp} and compared to the nominal
prediction for a perfectly saturated coating. The uncertainty on the points is
given by the standard deviation of the individual measurements and a 2\% fit
uncertainty added in quadrature.  
The coating procedure was found to reliably improve the QE of each individual
PMT close to the idealized nominal prediction with minimal deviations from the
average.

\begin{figure}
\centering
\includegraphics[width=1\linewidth]{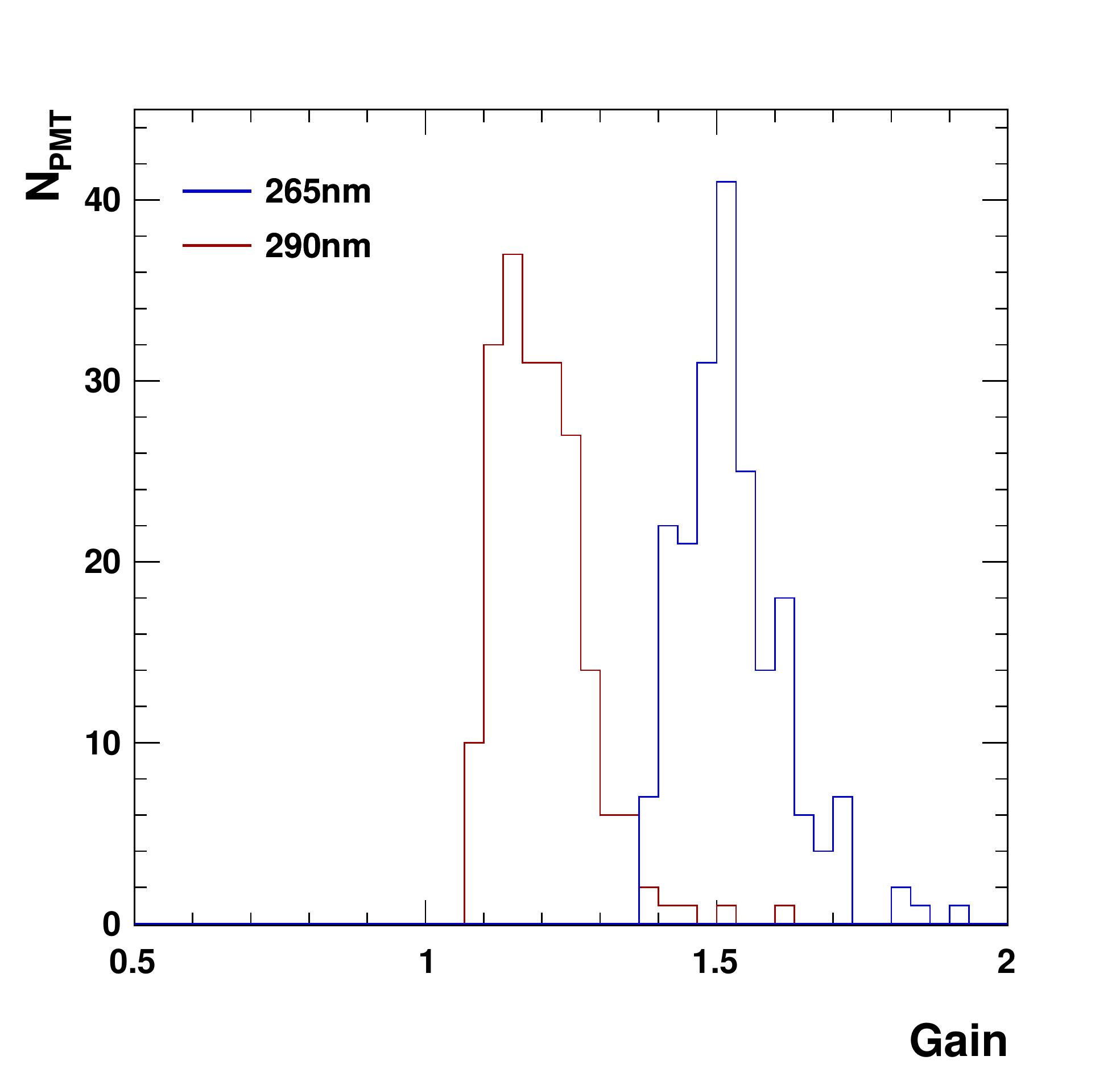}
\caption{
The measured gain for all 200 coated XP4500B PMTs for the 2-LED
rapid QA setup is shown on the $x$-axis, and the number of PMTs with this gain
value on the $y$-axis.
The blue histogram corresponds to the $265\,\text{nm}$
LED,
and the red histogram to the $290\,\text{nm}$ LED.
\label{fig:qa-hist}}
\end{figure}

The full spread of the measured gains for all 200 coated XP4500B PMTs is shown
in Fig.~\ref{fig:qa-hist}. The histograms are slightly asymmetric with a heavy
right tail, as the QA procedure rejected coated PMTs that showed insufficient
improvement.

The QA procedure was instrumental in the optimization
of the cleaning procedure of the PMT face before the vacuum evaporation.  
Once the this cleaning procedure was fully
established, the rejection rate was 5\%. The rejected PMTs were subsequently
carefully cleaned and re-coated, after which they passed the QA testing.

\subsection{Pulse shape and timing}

\begin{figure}
\centering
\includegraphics[width=1\linewidth]{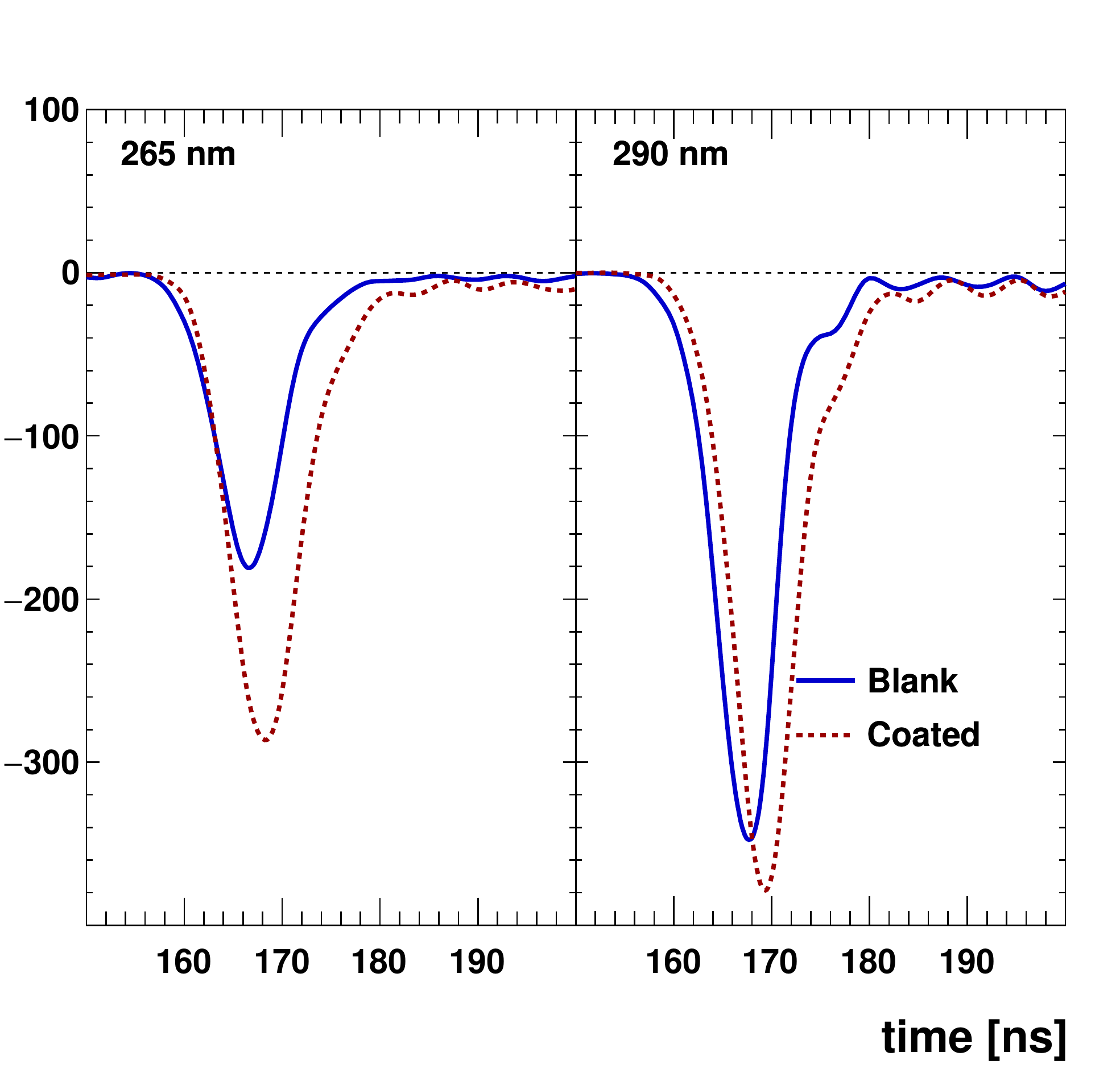}
\caption{
  Comparison of the average PMT response before (solid blue line) and after
  (dashed red line) coating. The coated PMT response is larger in amplitude, and
  delayed by approximately $2\,\text{ns}$. The left panel shows the results for
  the $265\,\text{nm}$ LED, and the right panel for the $290\,\text{nm}$ LED.
\label{fig:pulse}}
\end{figure}

The rapid QA setup was used to study the impact of the PT coating on the PMT
pulse shape and timing. The function generator provides a
consistent trigger when comparing PMTs before and after coating, while the 
CAEN V1729A FADC has a
timing resolution of $0.5\,\text{ns}$. 
Fig.~\ref{fig:pulse} shows a comparison
of the measured average pulse before and after coating. After coating, the pulse
is unchanged except for an increase in amplitude due to the higher PMT
efficiency and a delay of $2.0\pm0.5\,\text{ns}$. This observed
delay is consistent with an earlier measurement of the PT lifetime
\cite{Garwin:1973ex}.

\section{Summary}

Application of a p-terphenyl wavelength shifter to
the window of a UV-glass PMT through vacuum evaporation  was found to dramatically improve the photon
detection efficiency for wavelengths below $300\,\text{nm}$.
The wavelength dependent gain for Photons XP4500B PMTs was precisely measured
using a monochromator.

The ideal material load was found to be $110\pm10\,\mu\text{g}/\text{cm}^2$ with
a gain factor of 5.4 $\pm$ 0.5 at a wavelength of $215\,\text{nm}$.
Smaller material loads are not fully saturated while larger material loads are
more prone to a lower optical transparency due to crystallisation of the PT.

200 Photonis XP4500B PMTs were treated for the CLAS12 LTCC. Each PMT was tested
using a rapid QA procedure using two narrow-band UVTOP LEDs.
The coated PMTs will provide a 40\% increase in detector efficiency.

This work is supported in part by the U.S. Department of Energy Grant Award
DE-FG02-94ER4084.

\bibliography{WLS.bib}

\begin{thebibliography}{15}
\providecommand{\natexlab}[1]{#1}
\providecommand{\url}[1]{\texttt{#1}}
\expandafter\ifx\csname urlstyle\endcsname\relax
  \providecommand{\doi}[1]{doi: #1}\else
  \providecommand{\doi}{doi: \begingroup \urlstyle{rm}\Url}\fi

\bibitem[Adams and {others}(2001)]{ltcc:2001nim}
G.~Adams and {others}.
\newblock {The CLAS Cerenkov Detector}.
\newblock Nucl. Inst. and Meth. A 465:\penalty0 465, 2001.

\bibitem[Collaboration(2008)]{CLAS12:2008tr}
The~CLAS12 Collaboration.
\newblock {CLAS12 Technical Design Report}.
\newblock Technical report, JLab, Newport News, Virginia, July 2008.
\newblock URL \url{https://www.jlab.org/Hall-B/clas12_tdr.pdf}.

\bibitem[Leo(1987)]{Leo:1987kd}
W.~R. Leo.
\newblock \emph{{Techniques for Nuclear and Particle Physics Experiments: A How
  to Approach}}.
\newblock 1987.

\bibitem[{Photonis}(2007)]{Photonis:2007ta}
{Photonis}.
\newblock {Photomultiplier Tubes Catalogue}, January 2007.

\bibitem[Mai and Drouin(1971)]{Mai:71}
T~T~H Mai and R~Drouin.
\newblock {Relative Quantum Efficiencies of Some Ultraviolet Scintillators}.
\newblock \emph{Appl. Opt.}, 10\penalty0 (1):\penalty0 207--208, Jan 1971.
\newblock \doi{10.1364/AO.10.0207_1}.
\newblock URL \url{http://ao.osa.org/abstract.cfm?URI=ao-10-1-207_1}.

\bibitem[Garwin et~al.(1973)Garwin, Tomkiewicz, and Trines]{Garwin:1973ex}
E~L Garwin, Y~Tomkiewicz, and D~Trines.
\newblock {Method for elimination of quartz-face phototubes in Cherenkov
  counters by use of wavelength-shifter}.
\newblock \emph{Nuclear Instruments and Methods}, 107\penalty0 (2):\penalty0
  365--370, March 1973.

\bibitem[Alves et~al.(1974)Alves, Dos~Santos, and Policarpo]{Alves:1974em}
M~A~F Alves, M~C~M Dos~Santos, and A~J P~L Policarpo.
\newblock {Wavelength shifters for xenon proportional scintillation counters}.
\newblock \emph{Nuclear Instruments and Methods}, 119:\penalty0 405--406, July
  1974.

\bibitem[Koczon et~al.(2010)Koczon, Braem, Joram, Solevi, D{\"u}rr, and
  H{\"o}hne]{Koczon:1457653}
P~Koczon, A~Braem, C~Joram, P~Solevi, M~D{\"u}rr, and C~H{\"o}hne.
\newblock {Wavelength-shifting materials for the use in RICH detectors -
  p-terphenyl and tetraphenyl-butadiene revisited}.
\newblock Technical Report PH-EP-Tech-Note-2012-003, CERN, Geneva, Mar 2010.
\newblock URL \url{https://cds.cern.ch/record/1457653}.

\bibitem[Baillon et~al.(1975)Baillon, D{\'e}clais, Ferro-Luzzi, French, Jenni,
  Perreau, S{\'e}guinot, and Ypsilantis]{Baillon:1975jx}
P~Baillon, Y~D{\'e}clais, M~Ferro-Luzzi, B~French, P~Jenni, J~M Perreau,
  J~S{\'e}guinot, and T~Ypsilantis.
\newblock {Ultraviolet Cherenkov light detector}.
\newblock \emph{Nuclear Instruments and Methods}, 126\penalty0 (1):\penalty0
  13--23, May 1975.

\bibitem[Eigen and Lorenz(1979)]{Eigen:1979ev}
G~Eigen and E~Lorenz.
\newblock {A method of coating photomultipliers with wavelength shifters}.
\newblock \emph{Nuclear Instruments and Methods}, 166\penalty0 (2):\penalty0
  165--168, November 1979.

\bibitem[Grande and Moss(1983)]{Grande1983539}
Manuel Grande and Gary~R Moss.
\newblock {An optimised thin film wavelength shifting coating for Cherenkov
  detection}.
\newblock \emph{Nuclear Instruments and Methods in Physics Research},
  215\penalty0 (3):\penalty0 539--548, 1983.

\bibitem[Gorin et~al.(1986)Gorin, Kakauridze, Peresypkin, Polyakov, Rykalin,
  and Tzhadadze]{Gorin:1986ej}
A~M Gorin, G~D Kakauridze, A~I Peresypkin, V~A Polyakov, V~I Rykalin, and E~G
  Tzhadadze.
\newblock {On the increase of ultraviolet radiation detection efficiency in
  nuclear particle detectors with the help of transparent wavelength shifter
  films}.
\newblock \emph{Nuclear Inst. and Methods in Physics Research, A}, 251\penalty0
  (3):\penalty0 461--468, November 1986.

\bibitem[Dixon et~al.(2005)Dixon, Taniguchi, and Lindsey]{Dixon:2005cp}
James~M Dixon, Masahiko Taniguchi, and Jonathan~S Lindsey.
\newblock {PhotochemCAD 2: A Refined Program with Accompanying Spectral
  Databases for Photochemical Calculations}.
\newblock \emph{Photochemistry and Photobiology}, 81\penalty0 (1):\penalty0
  212--213, January 2005.

\bibitem[{Newport}(2010)]{Newport::CS260-USB-1-FH-A}
{Newport}.
\newblock {Monochromator, UV-VIS High Resolution, USB Interface, Fixed Slits},
  January 2010.

\bibitem[Bellamy et~al.(1994)Bellamy, Bellettini, Gervelli, Incagli, Lucchesi,
  and {others}]{Bellamy:1994bv}
E~H Bellamy, G~Bellettini, F~Gervelli, M~Incagli, D~Lucchesi, and {others}.
\newblock {Absolute calibration and monitoring of a spectrometric channel using
  a photomultiplier}.
\newblock \emph{Nucl.Instrum.Meth.}, A339:\penalty0 468--476, 1994.

\end{thebibliography}

\end{document}